\documentclass[12pt]{article}

\usepackage[super]{natbib} 
\usepackage{color} 
\usepackage[table]{xcolor} 
\usepackage{paralist}
\usepackage{geometry}
\usepackage{graphicx,graphics}
\usepackage{amsmath}	
\usepackage{amssymb}	
\usepackage{newtxtext,newtxmath}
\usepackage[T1]{fontenc}
\usepackage{ae,aecompl}
\usepackage{epsfig}   
\usepackage{microtype}
\usepackage[utf8]{inputenc}
\usepackage{enumitem}
\usepackage{ragged2e}

\def\arcsec{\hbox{$^{\prime\prime}$}}
\usepackage{times}
\newlist{thematic}{itemize}{8}
\setlist[thematic]{label=$\square$}
\usepackage{pifont}
\usepackage{enumitem} 
\usepackage{hyperref} 
\hypersetup{
    colorlinks,
    linkcolor={blue!80!black},
    citecolor={blue!80!black},
    urlcolor={blue!80!black}
}

\newcommand{\cmark}{\ding{51}}%
\newcommand{\done}{\rlap{$\square$}{\raisebox{2pt}{\large\hspace{1pt}\cmark}}%
\hspace{-2.5pt}}

\def\araa{{ARA\&A}}          
\def\apj{{ApJ}}                 
\def\apjl{{ApJ}}

\def\aap{ {A\&A}}

\def\mnras{ {MNRAS}}

\def\pasj{ {PASJ}}

\newcommand{\be}{\begin{equation}}
\newcommand{\ee}{\end{equation}}

\newcommand{\gtsima}{$\; \buildrel > \over \sim \;$}
\newcommand{\ltsima}{$\; \buildrel < \over \sim \;$}
\newcommand{\prosima}{$\; \buildrel \propto \over \sim \;$}
\newcommand{\gsim}{\lower.5ex\hbox{\gtsima}}
\newcommand{\lsim}{\lower.5ex\hbox{\ltsima}}
\newcommand{\simgt}{\lower.5ex\hbox{\gtsima}}
\newcommand{\simlt}{\lower.5ex\hbox{\ltsima}}
\newcommand{\simpr}{\lower.5ex\hbox{\prosima}}

\begin{document}


\begin{flushleft}
\huge 
Astro2020 Science White Paper \linebreak

\begin{center}
Supernova Remnants in High Definition
\end{center}

\normalsize

\noindent \textbf{Thematic Areas:} \hspace*{60pt} $\square$ Planetary Systems \hspace*{10pt} $\square$ Star and Planet Formation \hspace*{20pt}\linebreak
$\done$ Formation and Evolution of Compact Objects \hspace*{31pt} $\square$ Cosmology and Fundamental Physics \linebreak
  $\done$  Stars and Stellar Evolution \hspace*{1pt} $\square$ Resolved Stellar Populations and their Environments \hspace*{40pt} \linebreak
  $\square$    Galaxy Evolution   \hspace*{45pt} $\square$             Multi-Messenger Astronomy and Astrophysics \hspace*{65pt} \linebreak
  
\textbf{Principal Author:}

Name: Laura A. Lopez
 \linebreak						
Institution: The Ohio State University
 \linebreak
Email: lopez.513@osu.edu
 \linebreak
Phone: 814-404-1810
 \linebreak
 
\textbf{Co-authors:} 
  \linebreak
Brian J. Williams (NASA Goddard Space Flight Center), Samar Safi-Harb (University of Manitoba), Sangwook Park (University of Texas Arlington), Paul P. Plucinsky (Harvard-Smithsonian Center for Astrophysics), David Pooley (Trinity University), Tea Temim (Space Telescope Science Institute), Katie Auchettl (DARK Cosmology Centre, Niels Bohr Institute), Aya Bamba (University of Tokyo), Carles Badenes (University of Pittsburgh), Daniel Castro (Harvard-Smithsonian Center for Astrophysics), Kristen Garofali (University of Arkansas), Denis Leahy (University of Calgary), Patrick Slane (Harvard-Smithsonian Center for Astrophysics), Jacco Vink (University of Amsterdam), Benjamin F. Williams (University of Washington), J. Craig Wheeler (University of Texas Austin) \\
  
\end{flushleft}

\noindent
\textbf{Abstract}: Supernova remnants (SNRs) offer the means to study SN explosions, dynamics, and shocks at sub-parsec scales. X-ray observations probe the hot metals synthesized in the explosion and the TeV electrons accelerated by the shocks, and thus they are key to test recent, high-fidelity three-dimensional SN simulations. In this white paper, we discuss the major advances possible with X-ray spectro-imaging at arcsecond scales, with a few eV spectral resolution and a large effective area. These capabilities would revolutionize SN science, offering a three-dimensional view of metals synthesized in explosions and enabling population studies of SNRs in Local Group galaxies. Moreover, this future X-ray mission could detect faint, narrow synchrotron filaments and shock precursors that will constrain the diffusive shock acceleration process.

\clearpage

\noindent
{\bf I. The X-Ray View of Supernova Remnants} \\
\vspace{-12mm}
\begin{center}
\rule{\textwidth}{0.2mm}
\end{center}
\vspace{-5mm}
\noindent 


Supernovae (SNe) play an essential role in the Universe. Metals synthesized during the explosion chemically enrich galaxies, supplying fodder for dust and the next generation of stars. Their shock waves plow through the interstellar medium (ISM) for thousands of years, accelerating particles to extreme energies ($\sim$10$^{15}$~eV) and amplifying magnetic fields up to a thousand times that of the ISM. The shocks also heat surrounding gas and impart momentum, altering the phase structure of the ISM, shaping galaxies, and driving kpc-scale galactic winds.

~~~~ Although hundreds of supernovae (SNe) are found each year at optical wavelengths by dedicated surveys, they are often too distant to resolve the SN ejecta and the immediate surroundings of the exploded stars. Studies of the closest SNe, such as SN 1987A (McCray \& Fransson 2016), have advanced the field tremendously, but our understanding of SN progenitors and explosion mechanisms is hampered by the infrequency of nearby events.

~~~~ Supernova remnants (SNRs) offer the means to study SN explosions, dynamics, and shocks at sub-pc scales, and they are an important tool to explore the relationship between compact objects and their explosive origins. Observations of SNR morphologies, kinematics, and chemical abundances are crucial to test and constrain recent, high-fidelity 3D SN simulations. Metals synthesized in the explosions are shock-heated to $\sim$10$^{7}$~K temperatures (see Fig.~1), and TeV electrons accelerated by the forward shock emit synchrotron radiation at X-ray energies. Thus, X-ray observations are a crucial means to probe the bulk of SNR ejecta material and the particle acceleration process.

\begin{figure}[b!]
  \begin{center}
  \vskip-15pt
    \begin{minipage}[b]{0.74\linewidth}
\raisebox{0.5cm}{\epsfig{file=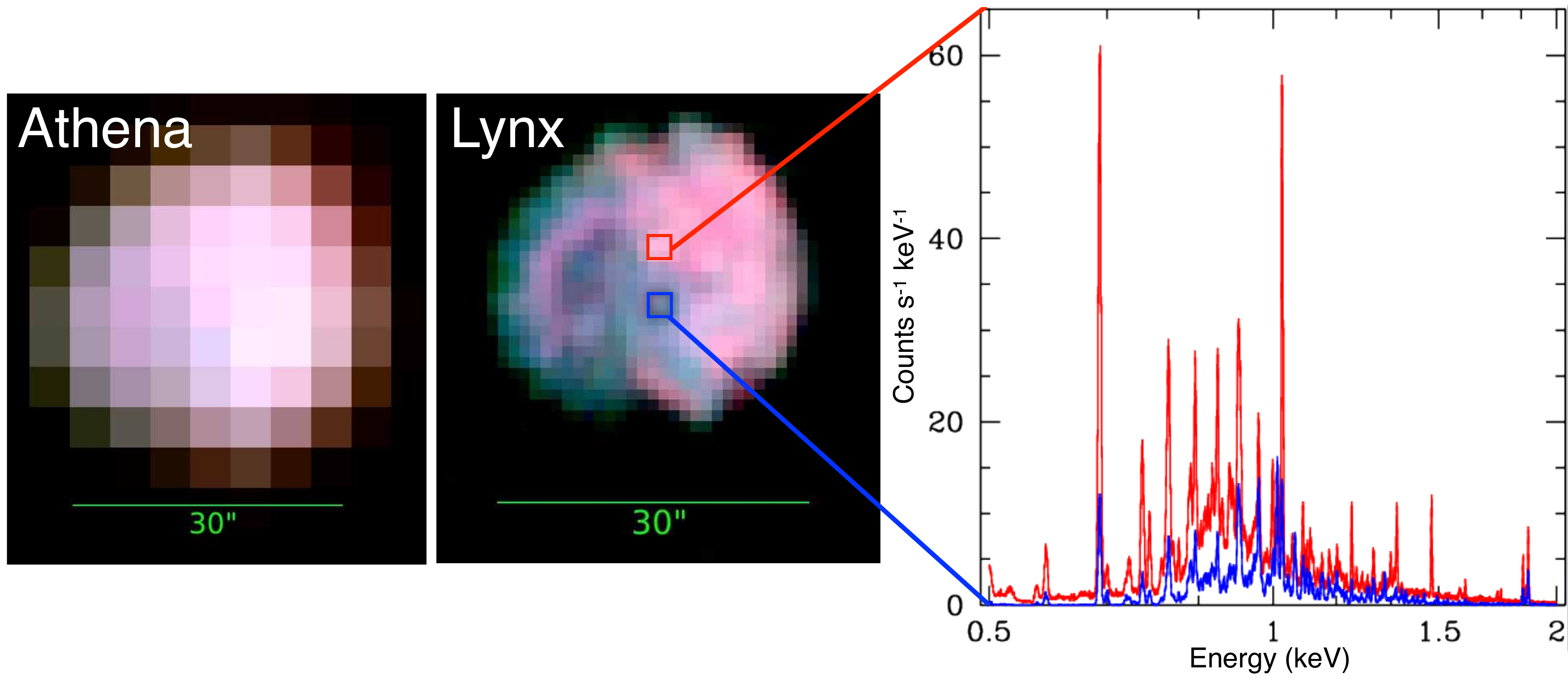, width=\linewidth}}
    \end{minipage}\hfill
    \begin{minipage}[b]{0.24\linewidth}
          \caption{\small Simulated {\it Athena} (left) and {\it Lynx} (middle) images of the Type Ia SNR N103B in the LMC. The right panel shows simulated microcalorimeter spectra from two locations in N103B. A 1\arcsec-pixel microcalorimeter is vital to obtain distinct spectra from the ejecta and the CSM components.}  
    \end{minipage}
  \end{center}
\end{figure}

~~~~ Future X-ray facilities offer exciting prospects for major advancements in SN science. Increased effective area relative to {\it Chandra} and {\it XMM-Newton} will enable detailed investigation of faint and distant SNRs, including the $>$600 SNRs in the Milky Way and Local Group galaxies (e.g., Badenes et al. 2010; Ferrand \& Safi-Harb 2012; Sasaki et al. 2012; Maggi et al. 2016; Garofali et al. 2017; Green 2017). Sub-arcsecond spatial resolution will enable proper motion studies over a baseline of several decades, and it will facilitate a resolved view of the thin synchrotron filaments around the periphery of young SNRs. X-ray microcalorimeters will resolve He-like and H-like line complexes of many elements, facilitating 3D mapping of metals synthesized in the explosions. With these capabilities, the sample size of young SNRs with morphological, kinematic, and nucleosynthetic measurements dramatically increases, and these observations are crucial to inform SN models and probe SN feedback and chemical enrichment in different Galactic environments. 

\begin{figure*}[t]
\begin{center}
\includegraphics[width=0.82\textwidth]{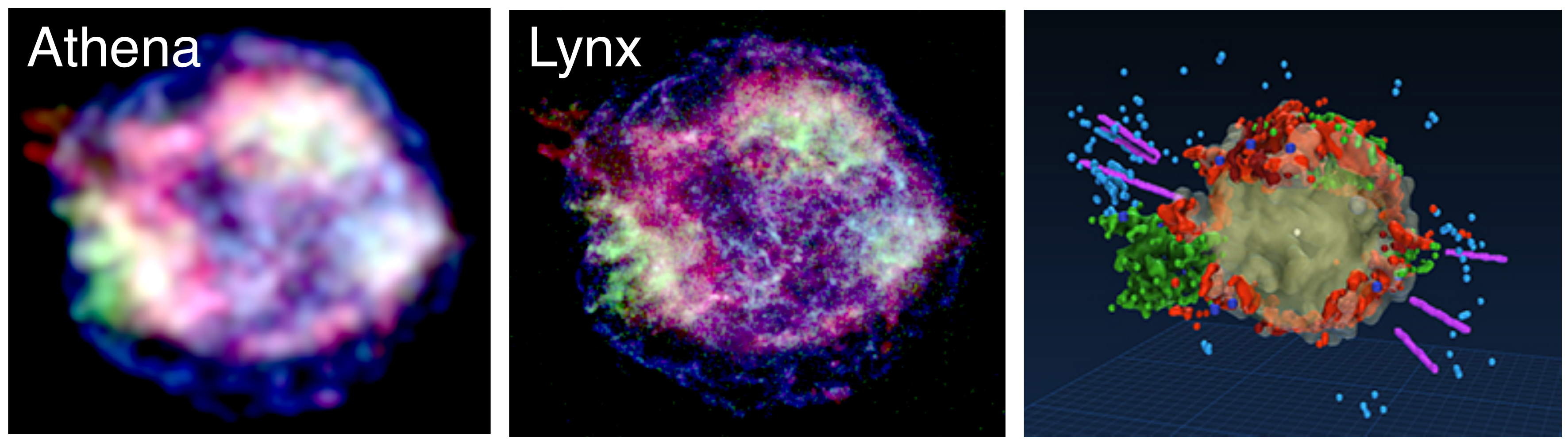}
\end{center}
 \vskip-20pt
\caption{\small Images of Cas~A, with 5\arcsec\ resolution (left) and 1\arcsec\ resolution (middle), corresponding to the spatial resolution of the proposed microcalorimeters on {\it Athena} and {\it Lynx}. {\it Right}:  3D view of Cas A, with Fe (X-rays) in green, Ar and Si (optical) in purple and blue, and cold ejecta (IR) in red. An X-ray microcalorimeter with 1\arcsec-pixels would yield a resolved 3D view of Cas A's hot ejecta, the vast majority of the metals synthesized in the SN explosion.}
 \label{fig:sizes}
\vskip-10pt
\end{figure*}

~~~~ In this white paper, we focus on the SNR science enabled by a future X-ray facility with sub-arcsecond spatial resolution capabilities (0.5\arcsec\ imaging), superb spectral resolution with spectro-imaging capabilities (a few eV at 1\arcsec\ scales), and a large effective area (2~m$^{2}$). In additional white papers, B. Williams et al. describes the major advances X-ray microcalorimeters will enable in understanding SN explosions, and S. Safi-Harb et al. summarizes the significant progress that spatial resolution and increased effective area provides regarding neutron stars (NSs) and pulsar wind nebulae (PWNe). 

\vskip5pt

\noindent
{\bf II. Resolving Galactic and Extragalactic SNRs} \\
\vspace{-12mm}
\begin{center}
\rule{\textwidth}{0.2mm}
\end{center}
\vspace{-3mm}
\noindent 

Since {\it Chandra}'s first-light image of Cassiopeia~A (Cas A) showing narrow, non-thermal filaments, small ejecta knots, and a neutron star at its center, the scientific benefit and beauty of high spatial resolution at X-ray energies became evident. However, a prime limitation of current facilities is that CCD energy resolution is insufficient to resolve He-like and H-like line complexes, and gratings spectrometers are only useful if SNRs have minimal angular extents (e.g., Dewey et al. 2008) or to study isolated ejecta knots (e.g., Bhalerao et al. 2015). 

~~~~ X-ray microcalorimeters (which are non-dispersive) will revolutionize SNR studies, and the few eV spectral resolution across these objects will yield the kinematics and a three-dimensional mapping of hot ejecta metals. The {\it Hitomi} spectra of the young SNR N132D, which showed bulk redshifted iron indicative of a highly asymmetric explosion (Hitomi Collaboration et al. 2018), gave a tantalizing glimpse of the power of high-resolution spectro-imaging capabilities. Within the Milky Way, the {\it Athena} microcalorimeter (with 5\arcsec\ pixels) will obtain superb spectra that will enable characterization of the individual components, e.g. the Si-rich jet, the Fe ahead of the forward shock, and the synchrotron filaments around the periphery (see Figure~2). 

With a 1\arcsec-pixel microcalorimeter, these features are resolved in even more detail, reducing confusion and enabling precise measures of the kinematics, shock heating, ejecta mixing, chemistry, and the non-thermal radiation. In particular, with a few eV spectral resolution on 1\arcsec\ scales, it is possible to obtain accurate radial velocities of ejecta knots in young, ejecta-dominated SNRs. Given the limitations associated with dispersed spectra from extended objects, current studies have only measured radial velocities for the brightest knots in a handful of SNRs (e.g., Bhalero et al. 2015). Thus, {\bf the combination of a few eV spectral resolution and 1\arcsec\ spatial resolution in the un-dispersed spectrum would truly revolutionize SNR studies, expanding the view of SNRs from two dimensions into 3D.} For example, Figure~2 shows the 3D view of Cas A, obtained using optical, infrared, and X-ray facilities. A microcalorimeter with 1\arcsec-pixels would yield a similar 3D map of young SNRs' hot ejecta, the majority of the metals synthesized in the SN explosion. In addition, improved effective area at hard X-ray energies ($\sim$6--8~keV, where current X-ray telescopes have low sensitivity) would facilitate constraints on Fe-K and the Fe-group elements, which are especially crucial to explore Type Ia progenitor scenarios.

\begin{figure*}[t]
\begin{center}
\includegraphics[width=0.85\textwidth]{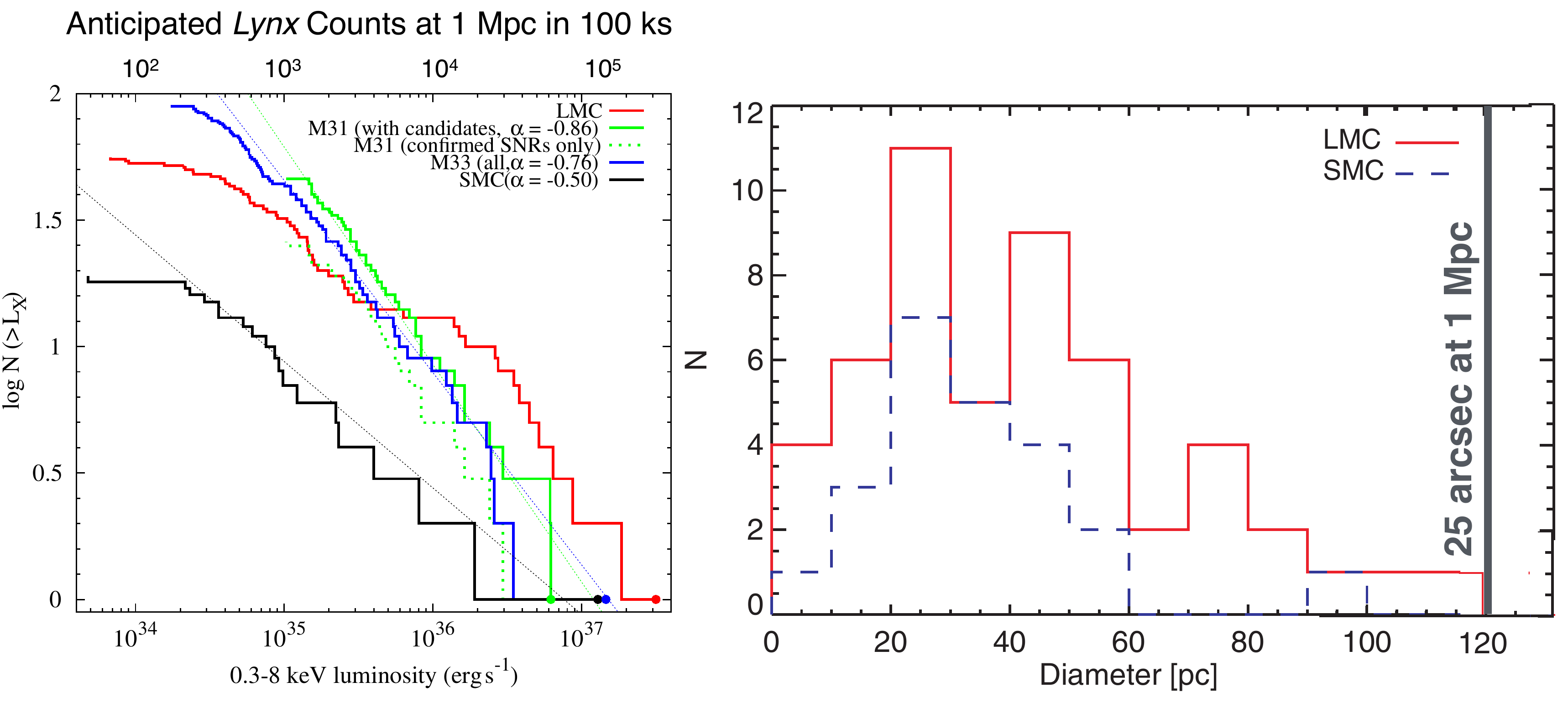}
\end{center}
 \vskip-22pt
\caption{\small {\it Left}: Luminosity function of known SNRs in nearby galaxies LMC, SMC, M31, and M33 (from Maggi et al. 2016). The anticipated number of counts in a 100~ks {\it Lynx} exposure from SNRs are shown along the top for a distance of 1~Mpc. {\it Right}: Histogram of the size distribution of SNRs in the LMC and SMC (Badenes et al. 2010). The dark grey line represents the physical size of 25\arcsec\ at a distance of 1~Mpc. Thus, arcsecond pixels would be sufficient to measure extension of the SNRs across many resolution elements, whereas 5\arcsec\ pixels would only resolve the largest SNRs.}
 \label{fig:sizes}
\vskip-10pt
\end{figure*}

~~~~ Spatial resolution is especially important to probe the morphologies and distinct spectral components in extragalactic SNRs. For example, Figure~1 shows simulated {\it Athena} and {\it Lynx} images of the young SNR N103B in the Large Magellanic Cloud (LMC). In this case, 1\arcsec\-pixels are crucial to disentangle the ejecta and circumstellar medium (CSM) spectra. At the distance of the LMC, 1\arcsec\ $\approx$~ 0.3~pc; thus with that resolution, maps of the youngest parsec-scale sources could be obtained, such as for SN~1987A (which will be 3\arcsec $\approx$~1~pc across in the 2030s: Orlando et al. 2015). 

~~~~ Improvements in effective area would enable detailed studies of large extragalactic populations (see Figure 3; left). For example, a 100-ks observation with a 2~m$^{2}$ effective area would detect $>$10$^{3}$ counts from the dozens of SNRs in M31 and M33. In the LMC and Small Magellanic Cloud (SMC), X-ray imaging and microcalorimeter observations could achieve similar signal-to-noise to what current facilities achieve for Milky Way sources. Furthermore, spatial resolution is crucial to resolve SNRs in the Local Group. For example, the known SNRs in M31 and M33 have angular sizes of $\sim$3--30\arcsec. Assuming Local Group SNRs have the same size distribution as observed in the LMC and SMC (Badenes et al. 2010), arcsecond spatial resolution is crucial to resolve SNRs across tens of pixels at a distance of 1~Mpc (see Figure 3, right). 

~~~~ As the morphologies and Fe-line centroid can be used to ``type" SNRs (Lopez et al. 2011; Yamaguchi et al. 2014), a sensitive X-ray telescope with arcsecond spatial resolution would foster constraints on the explosive origin of hundreds of SNRs in the Local Group. This sample would be large enough to do statistical comparisons of populations, in addition to studies of individual sources. These results would be useful to explore SN explosions in different Galactic environments and to probe SN feedback and enrichment in the Local Group. 

\clearpage

\noindent
{\bf III. Proper Motions Over Baselines of Several Decades} \\
\vspace{-12mm}
\begin{center}
\rule{\textwidth}{0.2mm}
\end{center}
\vspace{-3mm}
\noindent 

While a microcalorimeter will allow measurems of velocities along the line-of-sight, {\bf comparable spatial resolution to {\it Chandra} in a future X-ray facility would enable proper motion studies of many Galactic sources and SNR shocks over 30- to 40-year baselines.} For example, Table~1 shows the proper motions in a 30-year baseline for different distances ($D = $1, 5, 10, or 50~kpc) and velocities ($v = $100, 500, 1000, 5000~km s$^{-1}$). Depending on the object, proper motions can, in some cases, be measured better than the on-axis point spread function (PSF): e.g. Xi et al. (2019) used {\it Chandra} and measured an expansion of 0.1\arcsec\ over 17 years for the SNR 1E~0102.2$-$7219 (corresponding to a $v = 1600$ km~s$^{-1}$ for a distance of $D = 60$~kpc). Thus, velocities of 500 km~s$^{-1}$ may be measurable out to $D = 10$~kpc, and velocities of 1000 km~s$^{-1}$ may be observable out to the Magellanic Clouds. For high-velocity shocks of 5000~km~s$^{-1}$, sub-arcsecond angular resolution is sufficient to measure proper motions at the distance of the LMC, $D = 50$~kpc.

Given that the mean velocity of neutron stars (NSs) is 380~km~s$^{-1}$ (Faucher-Gigu{\`e}re \& Kaspi 2006), sub-arcsecond resolution over a 30- to 40-year baseline is sufficient to observe NS proper motions. As described in a separate white paper by S. Safi-Harb et al., these measurements set important constraints on the origin of NS kicks and on SN explosion models.

Combined with radial velocities from the microcalorimeter, proper motions will reveal the true 3D structure of the ejecta. Though {\it Chandra} has enabled proper motion measurements for a few SNRs (e.g., Yamaguchi et al. 2016), the off-axis PSF and sensitivity has limited the sample and the statistical significance of the results. Thus, an improved effective area and an arcsecond off-axis PSF would facilitate precise measurements of SNR dynamics across a much larger population.

\begin{table}[h]
\vskip-13pt
\begin{center}
\footnotesize
\caption{Proper Motions Over a 30-Year Baseline for Different Distances}
\begin{tabular}{lcccc}
\hline
Velocity & Proper Motion & Proper Motion & Proper Motion & Proper Motion \\
(km~s$^{-1}$) & for $D =$ 1~kpc  & for $D =$ 5~kpc  & for $D =$ 10~kpc  & for $D =$ 50~kpc \\ 
\hline
100 & 0.63\arcsec & 0.13\arcsec & 0.06\arcsec & 0.01\arcsec \\
500 & 3.16\arcsec & 0.63\arcsec & 0.32\arcsec & 0.06\arcsec \\
1000 & 6.33\arcsec & 1.27\arcsec & 0.63\arcsec & 0.12\arcsec \\
5000 & 31.6\arcsec & 6.33\arcsec & 3.16\arcsec & 0.63\arcsec \\
\hline
\end{tabular}
\end{center}
\vskip-20pt
\end{table}

\noindent
{\bf IV. Non-Thermal Emission from Fast Shock Waves} \\
\vspace{-12mm}
\begin{center}
\rule{\textwidth}{0.2mm}
\end{center}
\vspace{-3mm}
\noindent 

Several young SNRs (e.g., SN 1006, Tycho, and RCW 86), emit synchrotron emission in narrow filaments around their periphery. This synchrotron emission results from a non-thermal population of electrons, accelerated to relativistic energies behind the shock, spiraling around an amplified post-shock magnetic field. There are many open questions in this process: under what conditions do shocks efficiently accelerate particles? How are magnetic fields amplified, and how high can this amplification go? Shocks are ubiquitous in astrophysics, and these questions are relevant in all sorts of environments, from colliding galaxy clusters to AGN jets to the Earth's bow shock created by the interaction with the solar wind. Of these, only SNRs offer the chance to study shock physics on length scales comparable to those on which these processes operate.

\begin{figure*}
\begin{center}
\includegraphics[width=\textwidth]{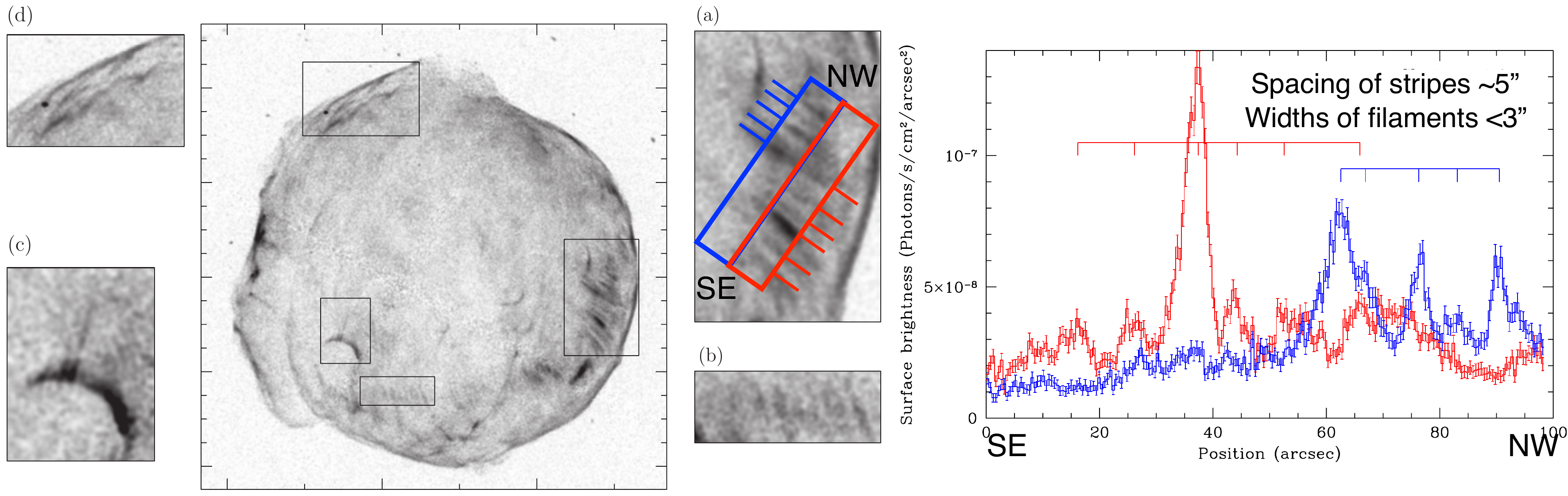}
\end{center}
 \vskip-23pt
\caption{\small {\it Left panels}: {\it Chandra} 4--6 keV image of the Tycho SNR from Eriksen et al. (2011). The non-thermal continuum dominates in this bandpass, and stripes and filaments are evident around the periphery and in projection inside the SNR. {\it Right}: Surface brightness profiles from the blue and red rectangles in panel a. The spacing of the stripes is 5\arcsec, and the width of the filaments is $<$3\arcsec. Thus, sub-arcsecond imaging is imperative to resolve these features, and a large effective area would enable the detection and characterization of the non-thermal filaments in SNRs.}
 \label{fig:filaments}
\vskip-10pt
\end{figure*}

~~~~ The width of the thin synchrotron rims ($\lsim$5\arcsec) in SNR  shocks (e.g., Bamba et al. 2005) offers a diagnostic of acceleration properties. In SN 1006, the width of these rims varies with energy, implying that the relativistic electrons rapidly age in a field as strong as 100 $\mu$G, inconsistent with the field damping quickly behind the shock (Ressler et al. 2014). However, in the well-studied Tycho SNR (shown in Figure~4), the thin synchrotron rims suggest strong amplification at the shock followed by quick damping (Tran et al. 2015). Thus, it is possible that two different mechanisms by which magnetic fields are amplified and subsequently damped in shock waves. 

~~~~ To obtain these results, the time investments from {\it Chandra} have been substantial (e.g., $\sim$1~Ms toward Tycho). An effective area of 2~m$^{2}$ and 1\arcsec\ resolution would ensure the detection of even faint non-thermal filaments and expand the sample of objects for which these measurements can be made. Combined with high-resolution radio data from the upgraded Very Large Array or the upcoming Square Kilometer Array, sensitive X-ray observations of the synchrotron continuum will yield crucial constraints on diffusive shock acceleration.
 
\begin{figure}[b]
  \begin{center}
  \vskip-20pt
    \begin{minipage}[b]{0.74\linewidth}
\raisebox{0cm}{\epsfig{file=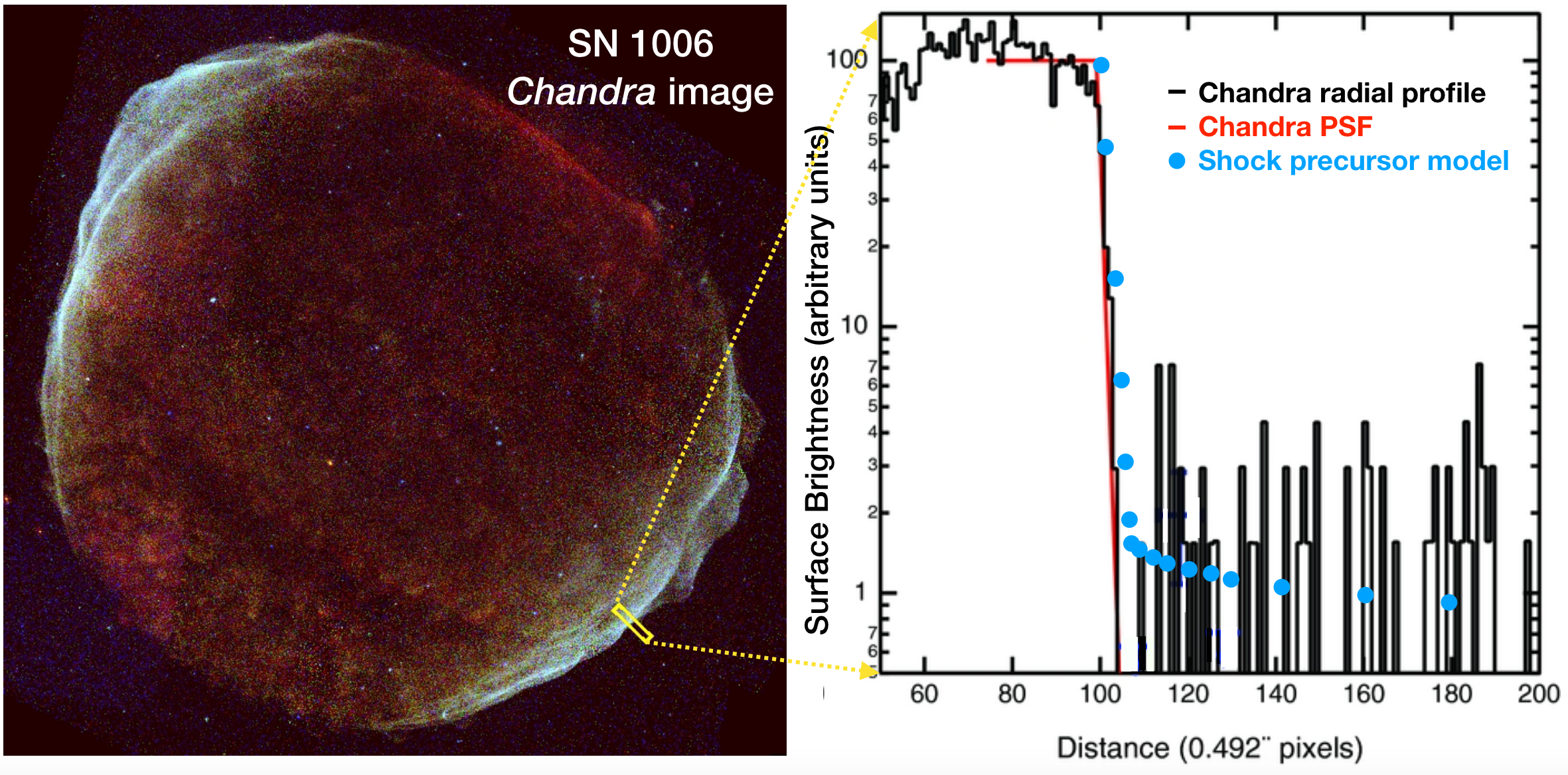, width=\linewidth}}
    \end{minipage}\hfill
    \begin{minipage}[b]{0.25\linewidth}
          \caption{\small {\it Chandra} image of SN 1006 (left), with a radial profile from the yellow box (right). The red curve shows the {\it Chandra} PSF; the blue dots show a potential shock precursor model (Morlino et al. 2010). Even with {\it Chandra}’s resolution, the background level is too high to detect whether a shock precursor (particles that have diffused upstream of the shock) is present.}  
    \end{minipage}
  \end{center}
\end{figure}

~~~~ Some accelerated particles must diffuse from behind the shock into the upstream medium. In SNR shocks dominated by non-thermal synchrotron emission from accelerated particles, faint X-ray emission should be present ahead of the shock, yet this emission has never been detected (Winkler et al. 2014; see Figure~5). Finding and characterizing this precursor will put tight constraints on the universal properties of fast shock waves, such as the degree of magnetic field amplification and the diffusion and scattering length of energetic particles. Arcsecond spatial resolution, a low background, and large effective area are critical to identifying these features in Galactic SNRs. \\

\noindent
{\bf References} \\
\vspace{-12mm}
\begin{center}
\rule{\textwidth}{0.2mm}
\end{center}
\vspace{-3mm}
\noindent 

\noindent Badenes, C., Maoz, D., \& Draine, B.~T.\ 2010, \mnras, 407, 1301

\noindent Bamba, A., Yamazaki, R., Yoshida, T., Terasawa, T., \& Koyama, K.\ 2005, \apj, 621, 793 

\noindent Bhalerao, J., Park, S., Dewey, D., et al.\ 2015, \apj, 800, 65 

\noindent Dewey, D., Zhekov, S.~A., McCray, R., \& Canizares, C.~R.\ 2008, \apjl, 676, L131 

\noindent Eriksen, K.~A., Hughes, J.~P., Badenes, C., et al.\ 2011, \apjl, 728, L28

\noindent Ferrand, G., \& Safi-Harb, S.\ 2012, Advances in Space Research, 49, 1313 

\noindent Garofali, K., Williams, B.~F., Plucinsky, P.~P., et al.\ 2017, \mnras, 472, 308 

\noindent Hitomi Collaboration, Aharonian, F., Akamatsu, H., et al.\ 2018, \pasj, 70, 16 

\noindent Lopez, L.~A., et al.\ 2011, \apj, 732, 114 

\noindent Maggi, P., Haberl, F., Kavanagh, P.~J., et al.\ 2016, \aap, 585, A162 

\noindent McCray, R., \& Fransson, C.\ 2016, \araa, 54, 19 

\noindent Morlino, G., Amato, E., Blasi, P. \& Caprioli, D. 2010, MNRAS, 405, 21

\noindent Orlando, S., Miceli, M., Pumo, M.~L., \& Bocchino, F.\ 2015, \apj, 810, 168 

\noindent Ressler, S.M., et al. 2014, ApJ, 790, 85

\noindent Sasaki, M., Pietsch, W., Haberl, F., et al.\ 2012, \aap, 544, A144 

\noindent Tran, A., Williams, B.J., Petre, R., Ressler, S.M, \& Reynolds, S.P. 2015, ApJ, 812, 101 

\noindent
Winkler, P.F., et al. 2014, ApJ, 781, 65

\noindent 
Xi, L., Gaetz, T.~J., Plucinsky, P.~P., Hughes, J.~P., \& Patnaude, D.~J.\ 2019, arXiv:1902.08456

\noindent 
Yamaguchi, H., Badenes, C., Petre, R., et al.\ 2014, \apjl, 785, L27 

\noindent Yamaguchi, H., Katsuda, S., Castro, D., et al.\ 2016, \apjl, 820, L3 

\end{document}